\documentclass[conference]{IEEEtran}
\IEEEoverridecommandlockouts

\usepackage{cite}
\usepackage{tabularx}
\usepackage{multirow}
\usepackage{amsmath,amssymb,amsfonts}
\usepackage{graphicx}
\usepackage[table,xcdraw]{xcolor}
\usepackage{algorithm}
\usepackage{algpseudocode}
\usepackage{bm}         
\usepackage{mathtools}  
\usepackage{adjustbox}
\usepackage{textcomp}
\usepackage{url}
\usepackage{float}
\usepackage{array}
\usepackage{cleveref}
\usepackage{comment}

\algnewcommand\algorithmicswitch{\textbf{switch}}
\algnewcommand\algorithmiccase{\textbf{case}}
\algnewcommand\algorithmicassert{\texttt{assert}}
\algnewcommand\Assert[1]{\State \algorithmicassert(#1)}

\newcommand{\eg}{{\em e.g.}}
\newcommand{\etal}{{\em et al.}}

\begin{document}
\title{CPINN-ABPI: Physics-Informed Neural Networks for Accurate Power Estimation in MPSoCs 
\vspace{-3mm}
}


\author{
    \IEEEauthorblockN{Mohamed R. Elshamy\IEEEauthorrefmark{1}, Mehdi Elahi\IEEEauthorrefmark{2}, Ahmad Patooghy\IEEEauthorrefmark{2}, and Abdel-Hameed A. Badawy\IEEEauthorrefmark{1}}
    \IEEEauthorblockA{\IEEEauthorrefmark{1}Klipsch School of ECE, New Mexico State University, Las Cruces, NM 88003, United States}
    \IEEEauthorblockA{\IEEEauthorrefmark{2}Computer Systems Technology, North Carolina A\&T State University, Greensboro, NC, United States}
    \IEEEauthorblockA{\IEEEauthorrefmark{1}\{elshamy, badawy\}@nmsu.edu, \IEEEauthorrefmark{2}melahi@aggies.ncat.edu, apatooghy@ncat.edu}
}

\maketitle

\begin{abstract}
Efficient thermal and power management in modern multiprocessor systems-on-chip (MPSoCs) demands accurate power consumption estimation. One of the state-of-the-art approaches, Alternative Blind Power Identification (ABPI), theoretically eliminates the dependence on steady-state temperatures, addressing a major shortcoming of previous approaches. 
However, ABPI performance has remained unverified in actual hardware implementations. In this study, we conduct the first empirical validation of ABPI on commercial hardware using the NVIDIA Jetson Xavier AGX platform. Our findings reveal that, while ABPI provides computational efficiency and independence from steady-state temperature, it exhibits considerable accuracy deficiencies in real-world scenarios. To overcome these limitations, we introduce a novel approach that integrates Custom Physics-Informed Neural Networks (CPINNs) with the underlying thermal model of ABPI. Our approach employs a specialized loss function that harmonizes physical principles with data-driven learning, complemented by multi-objective genetic algorithm optimization to balance estimation accuracy and computational cost. In experimental validation,
CPINN-ABPI achieves a reduction of $84.7\%$ CPU and $73.9\%$ GPU in the mean absolute error (MAE) relative to ABPI, with the weighted mean absolute percentage error (WMAPE) improving from $47\%$-$81\%$ to $\sim12\%$. The method maintains real-time performance with $195.3~\mu$s of inference time, with similar $85\%$-$99\%$ accuracy gains across heterogeneous SoCs.
\end{abstract}


\section{INTRODUCTION}
As chip complexity increases, power and thermal management have become critical aspects of modern multiprocessor system-on-chips (MPSoCs)~\cite{8714794,10678921}. To effectively manage power consumption while maximizing performance, MPSoCs require accurate, fine-grained power information with high temporal resolution (microseconds to milliseconds) and spatial resolution (unit level)~\cite{BPI_V1,Cluster-BPI}. Direct measurement of unit-level power is impractical due to the cost and complexity of per-unit sensors~\cite{BPI_V2}. Perfect electrical isolation is needed to accurately measure the power of individual units. However, the units in MPSoCs share power networks, making it nearly impossible to isolate the exact power consumption of each unit. Instead, modern MPSoCs rely on power models that estimate power consumption using observable performance counters, operating frequency, and voltage. Conventional approaches typically employ linear modeling techniques to establish the relationship between performance counters and dynamic power. Although these methodologies are computationally efficient, they encounter difficulties in accurately capturing the intricate and nonlinear relationships arising in contemporary architectures \eg, heterogeneous architectures exemplified by Snapdragon$_{TM}$ 8 Gen 3~\cite{snapdragon8gen3}. 

MPSoCs typically only have temperature sensors because they are much cheaper than power sensors. Since power and temperature are closely related, combining both data types helps achieve better energy efficiency and prevents overheating~\cite{BPI_V1}. Temperature serves as a late power consumption indicator, meaning that by the time a temperature sensor detects a hotspot, the unit has already been consuming excessive power for some time. Power information enables predictive thermal management rather than reactive, allowing systems to prevent thermal problems before they manifest and cause performance degradation or hardware damage~\cite{8474284}. 

Information about an MPSoC's power consumption is critical for performance optimization within established power constraints, enabling dynamic allocation of computational resources based on actual consumption rather than thermal effects. Furthermore, temperature does not directly correlate with energy efficiency in different operational scenarios, since a unit might maintain relatively low temperatures while simultaneously wasting energy on unnecessary operations or inefficient execution patterns. The challenge of thermal coupling adds another layer of complexity, where heat spreads between adjacent units, making it difficult to determine which specific unit actually generates heat based solely on temperature~\cite{10.1145/2566661}. Furthermore, fine-grained power estimation is very important nowadays for detecting thermal Trojans on hardware in MPSoCs~\cite{BIC,Matter,Cluster-BPI}. 

Power modeling for MPSoCs follows four main approaches. The first assumes a linear link between performance counters and power use, adjusting frequency and voltage by power-scaling laws~\cite{LinearModel1, LinearModel2}. Although efficient, it often misses complex power behaviors. The second uses higher-order terms for nonlinear dependencies, but as McCullough~\etal~\cite{Polynomial1} noted, it risks overfitting and poor generalization. The third employs feedforward neural networks for server-level energy accounting, which shows promising accuracy at coarse temporal resolutions~\cite{ANN1, ANN2, FFNN}. The fourth derives MPSoC models and unit power use from temperature and power measurements, providing rapid and precise estimates~\cite{BPI_V2, Cluster-BPI, ABPI}.

Previous approaches, such as Blind Power Identification (BPI)~\cite{BPI_V2}, BPI with Steady State (BPISS)~\cite{BPISS}, and Improved Cluster BPI (ICBPI)~\cite{Cluster-BPI} have attempted to estimate unit-level power consumption. However, these methods are dependent on steady-state temperature measurements, which are difficult to measure in practice. Alternative Blind Power Identification (ABPI) represents a significant advancement in eliminating dependence on steady-state temperatures, requiring only temperature measurements and total power consumption to estimate MPSoC parameters and the power for each unit~\cite{ABPI}. However, ABPI faces accuracy limitations due to simplifications in its physical model and interference from other MPSoCs components. To date, empirical validation of ABPI on an MPSoC against its embedded power sensors has not been carried out. Consequently, this study aims to perform the initial validation of ABPI utilizing an NVIDIA Jetson AGX Xavier board~\cite{nvidia_jetson_agx_xavier}. 
Our evaluations show that the ABPI estimates are inaccurate. 

We propose a new approach to enhance the precision of ABPI while maintaining its independence from steady-state temperature measurements. Our method leverages Physics-Informed Neural Networks (PINNs) with a custom loss function that balances the physical constraints derived from ABPI with empirical data fitting~\cite{PINN_OriginalPaper}. Furthermore, we employ the non-dominated sorting genetic algorithm II (NSGA-II) for multi-objective optimization of both accuracy and computational overhead, ensuring that the resulting model is suitable for real-time power management applications~\cite{NSGA-II_OriginalPaper, NSGA-PINN}. \\
This paper is organized as follows. Section~\ref{Proposed_App} presents the proposed CPINN-ABPI approach, including the neural network architecture and NSGA-II multiobjective optimization. Section~\ref{EXP_Setup} describes the experimental setup on NVIDIA Jetson AGX Xavier and simulated heterogeneous SoCs. Section~\ref{EXP_Results} presents the results comparing CPINN-ABPI against the traditional ABPI approach. Section~\ref{sec:conclusion} concludes the paper.

\section{Proposed Approach}
\label{Proposed_App}


Traditional ABPI suffers from significant limitations in its physics model, which negatively impacts the accuracy of its power estimations. As seen in the partial differential equation of heat diffusion~\cite{heatdiffusion}, Eqn.~\ref{equ:1}:

\begin{equation}
\rho(x, y, z) c_p(x, y, z) \frac{\partial t}{\partial \tau} = \nabla \cdot [\kappa(x, y, z) \nabla t] + p(x, y, z, \tau)
\label{equ:1}
\end{equation}

However, ABPI collapses this complex and continuous equation into a linearized, simplistic state-space model~\cite{BPI_V2} shown in Eqn.~\ref{equ:2}: 

\begin{equation}
T_r(k) = A T_r(k - 1) + B P(k)
\label{equ:2}
\end{equation} 
where the matrices $A$ and $B$ absorb both spatial coupling and temporal integration effects. This approximation implies the following key inaccuracies: \textbf{lost spatial detail}, \textbf{linearization \& constant properties}, \textbf{discretization error}, and \textbf{steady-state dependence}.


Our proposed approach (CPINN-ABPI) addresses ABPI limitations through a hybrid architecture combining physics-based modeling with neural networks via four key innovations: \textbf{1) A physics-informed residual network integration} that preserves ABPI's physical interpretability while improving accuracy through a custom neural architecture with a parallel residual branch; \textbf{2) A custom multi-component loss function} that enforces thermodynamic consistency and boundary conditions while learning complex empirical behaviors; \textbf{3) Transfer learning for thermal matrices} by initializing and fine-tuning ABPI's \(A\) and \(B\) matrices during training to adapt thermal parameters from measurements; and \textbf{4) Multi-objective optimization with NSGA-II} to jointly balance accuracy and computational efficiency through automatic hyperparameter selection. We discuss these innovations below.

\subsection{Custom Physics-Informed Neural Network (CPINN)}

CPINN-ABPI consists of two parallel branches:  a \textbf{physics branch} that implements the core ABPI algorithm and a \textbf{residual network branch} that learns to correct errors in the physics-based estimation. 








The \textbf{physics branch} implements the standard ABPI algorithm, which models the thermal behavior of the system using matrices $A$ and $B$: $\hat{\mathbf{t}} = \mathbf{t}_{\text{prev}} \mathbf{A}^T$ where $\hat{\mathbf{t}}$ represents the predicted temperature based on the previous temperature $\mathbf{t}_{\text{prev}}$ and the thermal conductivity matrix $\mathbf{A}$. The residual temperature is then calculated as $\Delta \mathbf{t} = \mathbf{t}_{\text{current}} - \hat{\mathbf{t}}$. Finally, the physics-based power estimation is derived using: $\mathbf{p}_{\text{physics}} = \Delta \mathbf{t} \cdot (\mathbf{B}^{-1})^T$ 
where $\mathbf{B}$ represents the thermal response matrix that maps power inputs to temperature changes. To incorporate prior knowledge and improve the convergence of the model, we initialize the physical matrices \( A \) and \( B \) in our CPINN-ABPI model using the final trained values obtained from the ABPI framework. This serves as a form of transfer learning, allowing the model to take advantage of previously learned thermals while adapting to new data and/or workloads. The model refines these physics-based parameters during training to better fit the current context. This strategy ensures a strong starting point grounded in physics, while enabling improved predictive accuracy through data-driven fine-tuning.









The \textbf{residual network branch} improves physics-based estimation by taking input $\mathbf{x} = [\mathbf{t}_{\text{prev}}, \mathbf{t}_{\text{current}}, \mathbf{P}_{\text{estimated}}]$ where $\mathbf{P}_{\text{estimated}}$ represents the output of the ABPI algorithm, which constitutes a variant of transfer learning. The network outputs a correction term $\Delta \mathbf{p} = \text{ResidualNetwork}(\mathbf{x})$ through a multilayer structure with fully connected layers and non-linear activation functions designed to capture complex relationships that the simplified physics model could miss. The final power estimation combines the physics-based estimate with the residual network correction through $\mathbf{p}_{\text{final}} = \mathbf{p}_{\text{physics}} + \Delta \mathbf{p}$. This hybrid approach leverages the robustness and interpretability of the physics-based approach while using the neural network to capture complex nonlinear relationships that the simplified physics model cannot represent. 

\subsection{Custom Multi-Component Loss Function}

A key contribution of our approach is the custom physics-informed loss function that guides the learning process toward solutions that respect physical constraints while improving accuracy. Our loss function consists of three components:

\subsubsection{Data Fidelity Loss}
This component measures the discrepancy between the predicted and actual power values as shown in Eqn.~\ref{eq:3}.

\begin{equation}
\mathcal{L}_{\text{data}} = \frac{1}{N} \sum_{i=1}^{N} \|\mathbf{p}_{\text{final}}^{(i)} - \mathbf{p}_{\text{true}}^{(i)}\|^2
\label{eq:3}
\end{equation} 
This term ensures that the model's predictions align with the ground truth measurements. 

\subsubsection{Physics Consistency Loss}
This component enforces consistency with the thermal dynamics equations as shown in Eqn.~\ref{eq:4}. 
\begin{equation}
\mathcal{L}_{\text{phys}} = \frac{1}{N} \sum_{i=1}^{N} \|\mathbf{t}_{\text{prev}}^{(i)}\mathbf{A}^T + \mathbf{p}_{\text{final}}^{(i)}\mathbf{B}^T - \mathbf{t}_{\text{current}}^{(i)}\|^2
\label{eq:4}
\end{equation} 
This term ensures that the predicted power values produce temperature changes consistent with the observed data when used in the thermal behavior equation. This is a critical aspect of the physics-informed approach, as it embeds the physics laws governing the system directly into the learning process. 

\subsubsection{Physics Guidance Loss:} This component gently steers the residual correction of the network toward the pure physics estimate as shown in Eqn.~\ref{eq:5}
\begin{equation}    
\mathcal{L}_{\mathrm{guide}}
= \frac{1}{N} \sum_{i=1}^N \bigl\lVert\,p_{\mathrm{final}}^{(i)} - p_{\mathrm{physics}}^{(i)}\bigr\rVert^2
\label{eq:5}
\end{equation} 
This term ensures that the learned correction does not stray too far from the prediction of analytical power, improves training stability, preserves physical interpretability, and prevents degenerate solutions where the network might ignore the underlying physics model.

The complete loss function combines these components with appropriate weighting as shown in Eqn.~\ref{eq:6} 
\begin{equation}
\mathcal{L}
= \mathcal{L}_{\mathrm{data}}
+ \lambda_{\mathrm{phys}}\,\mathcal{L}_{\mathrm{phys}}
+ \lambda_{\mathrm{guide}}\,\mathcal{L}_{\mathrm{guide}}
\label{eq:6}
\end{equation} 
where $\lambda_{\text{phys}}$,$\lambda_{\text{guide}}$  are hyperparameters that control the influence of the physics consistency and guide losses, respectively. These hyperparameters are automatically tuned through the NSGA-II multi-objective optimization process described below.

\subsection{Multi-Objective Optimization with NSGA-II}

Although increasing the complexity of the residual network can improve accuracy, it also increases computational overhead, which is undesirable for real-time power management applications. To address this tradeoff, we employ NSGA-II~\cite{NSGA-II_OriginalPaper, NSGA-PINN}, a multiobjective genetic algorithm, to optimize the model architecture and hyperparameters.

The optimization process explores various design choices, including the residual network architecture (number of layers, layer widths, activation functions), loss function weights (\(\lambda_{\mathrm{phys}}\), \(\lambda_{\mathrm{guide}}\)) and regularization parameters (dropout rates, weight decay)—and strives to balance two primary objectives: accuracy and overhead. Accuracy is measured by the mean absolute error (MAE) and the coefficient of determination (\(R^2\)), while computational overhead is quantified by the number of multiply–accumulate (MAC) operations.

To ensure the robustness and generalizability of the selected model architecture, we apply a 10-fold cross-validation on the best-performing configuration obtained from the genetic algorithm. This step helps to assess whether the model is overfitting and provides a more reliable estimate of its performance on unseen data. For each fold, the model is trained on 90\% of the data and evaluated on the remaining 10\%, cycling through all partitions. The evaluation is based on three standard metrics: Mean Squared Error (MSE), Mean Absolute Error (MAE), and the Coefficient of Determination (\(R^2\)). These metrics are computed for each output component \(j\), where \( p_i^{(j)} \) denotes the ground truth power value for the \(j^{th}\) component (\eg, CPU or GPU) at sample \(i\), and \( \hat{p}_i^{(j)} \) is the corresponding predicted power value from the model. In the \( R^2 \) calculations, \( \bar{p}^{(j)} \) represents the mean of the actual ground truth values for output \(j\) across all samples, serving as a baseline for comparison. 

    
    

The complete workflow of the proposed (CPINN-ABPI) approach is detailed in~\cref{alg:cpinn-abpi} and~\cref{alg:nsga2}.

\begin{algorithm}[tb]
\caption{CPINN-ABPI}
\label{alg:cpinn-abpi}
\begin{algorithmic}[1]
\Require Temperature sequences $\{T(k)\}_{k=1}^K \in \mathbb{R}^{N_c}$, power data $\{P_{\text{est}}(k), P_{\text{act}}(k)\}$
\Ensure Trained CPINN-ABPI model

\Statex \textbf{Phase 1: Initialization}
\State Load thermal matrices $\mathbf{A}, \mathbf{B} \in \mathbb{R}^{N_c \times N_c}$ from empirical ABPI
\State Initialize trainable parameters: $\mathbf{A}' \leftarrow \mathbf{A}$, $\mathbf{B}' \leftarrow \mathbf{B}$
\State Initialize neural network $f_\theta$ with given architecture

\Statex \textbf{Phase 2: Training Loop}
\For{epoch $e = 1$ to $E_{\max}$}
    \State $\text{total\_loss} \leftarrow 0$
    \For{each batch $b$ in training data}
        \For{each sample $(T_{\text{prev}}, T_{\text{curr}}, P_{\text{est}}, P_{\text{act}})$ in batch $b$}
            
            \Statex \quad \textit{Physics-informed forward pass}
            \State $\hat{T} \leftarrow \mathbf{A}'^T \cdot T_{\text{prev}}$
            \State $\Delta T \leftarrow T_{\text{curr}} - \hat{T}$
            \State $P_{\text{physics}} \leftarrow \Delta T \cdot (\mathbf{B}'^{-1})^T$
            
            \Statex \quad \textit{Neural network correction}
            \State $x \leftarrow [T_{\text{prev}}; T_{\text{curr}}; P_{\text{est}}]$
            \State $\Delta P \leftarrow f_\theta(x)$
            \State $P_{\text{final}} \leftarrow P_{\text{physics}} + \Delta P$
            
            \Statex \quad \textit{Loss computation}
            \State $\mathcal{L}_{\text{data}} \leftarrow \|P_{\text{final}} - P_{\text{act}}\|^2$
            \State $\mathcal{L}_{\text{physics}} \leftarrow \|\mathbf{A}'^T T_{\text{prev}} + \mathbf{B}'^T P_{\text{final}} - T_{\text{curr}}\|^2$
            \State $\mathcal{L}_{\text{guide}} \leftarrow \|P_{\text{final}} - P_{\text{physics}}\|^2$
            \State $\mathcal{L}_{\text{batch}} \leftarrow \mathcal{L}_{\text{data}} + \lambda_{\text{phys}} \mathcal{L}_{\text{physics}} + \lambda_{\text{guide}} \mathcal{L}_{\text{guide}}$
            \State $\text{total\_loss} \leftarrow \text{total\_loss} + \mathcal{L}_{\text{batch}}$
        \EndFor
    \EndFor
    
    \Statex \quad \textit{Backward pass and optimization}
    \State Compute gradients: $\nabla_\theta \text{total\_loss}$, $\nabla_{\mathbf{A}'} \text{total\_loss}$, $\nabla_{\mathbf{B}'} \text{total\_loss}$
    \State Update parameters: $\theta, \mathbf{A}', \mathbf{B}'$ using optimizer
    
    \Statex \quad \textit{Convergence check}
    \If{validation loss converged}
        \State \textbf{break}
    \EndIf
\EndFor

\State \Return Optimized model $(\theta^*, \mathbf{A}'^*, \mathbf{B}'^*)$
\end{algorithmic}
\end{algorithm}

\begin{algorithm}[tb]
\caption{NSGA-II for CPINN-ABPI Architecture Optimization}
\label{alg:nsga2}
\begin{algorithmic}[1]
\Require Training dataset $\mathcal{D}$, population size $N_{\text{pop}}$, generations $G_{\text{max}}$
\Ensure Pareto-optimal architectures $\mathcal{P}^*$

\Statex \textbf{Phase 1: Population Initialization}
\State Initialize population $P_0 \leftarrow \{\}$
\For{$i = 1$ to $N_{\text{pop}}$}
    \State Generate random individual $\mathbf{h}_i \leftarrow (\text{layers}, \text{widths}, \text{activation}, \lambda_{\text{phys}}, \lambda_{\text{guide}})$
    \State $P_0 \leftarrow P_0 \cup \{\mathbf{h}_i\}$
\EndFor

\Statex \textbf{Phase 2: Evolution Loop}
\For{generation $g = 0$ to $G_{\max} - 1$}
    
    \Statex \quad \textit{Objective evaluation}
    \For{each individual $\mathbf{h}_i \in P_g$}
        \State Train CPINN-ABPI model using Algorithm 1 with hyperparameters $\mathbf{h}_i$
        \State Evaluate MAE: $f_1(\mathbf{h}_i) \leftarrow \frac{1}{|\mathcal{D}_{\text{test}}|} \sum \|P_{\text{pred}} - P_{\text{true}}\|_1$
        \State Evaluate computational cost: $f_2(\mathbf{h}_i) \leftarrow \text{MACs}(\mathbf{h}_i)$
        \State Assign objectives: $\mathbf{f}(\mathbf{h}_i) \leftarrow [f_1(\mathbf{h}_i), f_2(\mathbf{h}_i)]$
    \EndFor
    
    \Statex \quad \textit{Non-dominated sorting}
    \State Compute domination relationships for all individuals in $P_g$
    \State Partition $P_g$ into fronts: $\mathcal{F}_1, \mathcal{F}_2, \ldots, \mathcal{F}_k$
    \State Compute crowding distance for each individual
    
    \Statex \quad \textit{Selection and reproduction}
    \If{$g < G_{\max} - 1$}
        \State Generate offspring $Q_g$ through tournament selection, crossover, and mutation
        \State Combine populations: $R_g \leftarrow P_g \cup Q_g$
        \State Apply elitist selection to form $P_{g+1}$ of size $N_{\text{pop}}$
    \EndIf
\EndFor

\Statex \textbf{Phase 3: Pareto Front Extraction}
\State Extract final Pareto front: $\mathcal{P}^* \leftarrow \mathcal{F}_1$ from $P_{G_{\max}}$
\State Apply the 10-Fold Cross-Validation on the Pareto solution.
\State \Return $\mathcal{P}^*$
\end{algorithmic}
\end{algorithm}

\section{Experimental Setup}
\label{EXP_Setup}
\begin{figure}[tb] 
\centering
\includegraphics[width=0.99\linewidth]{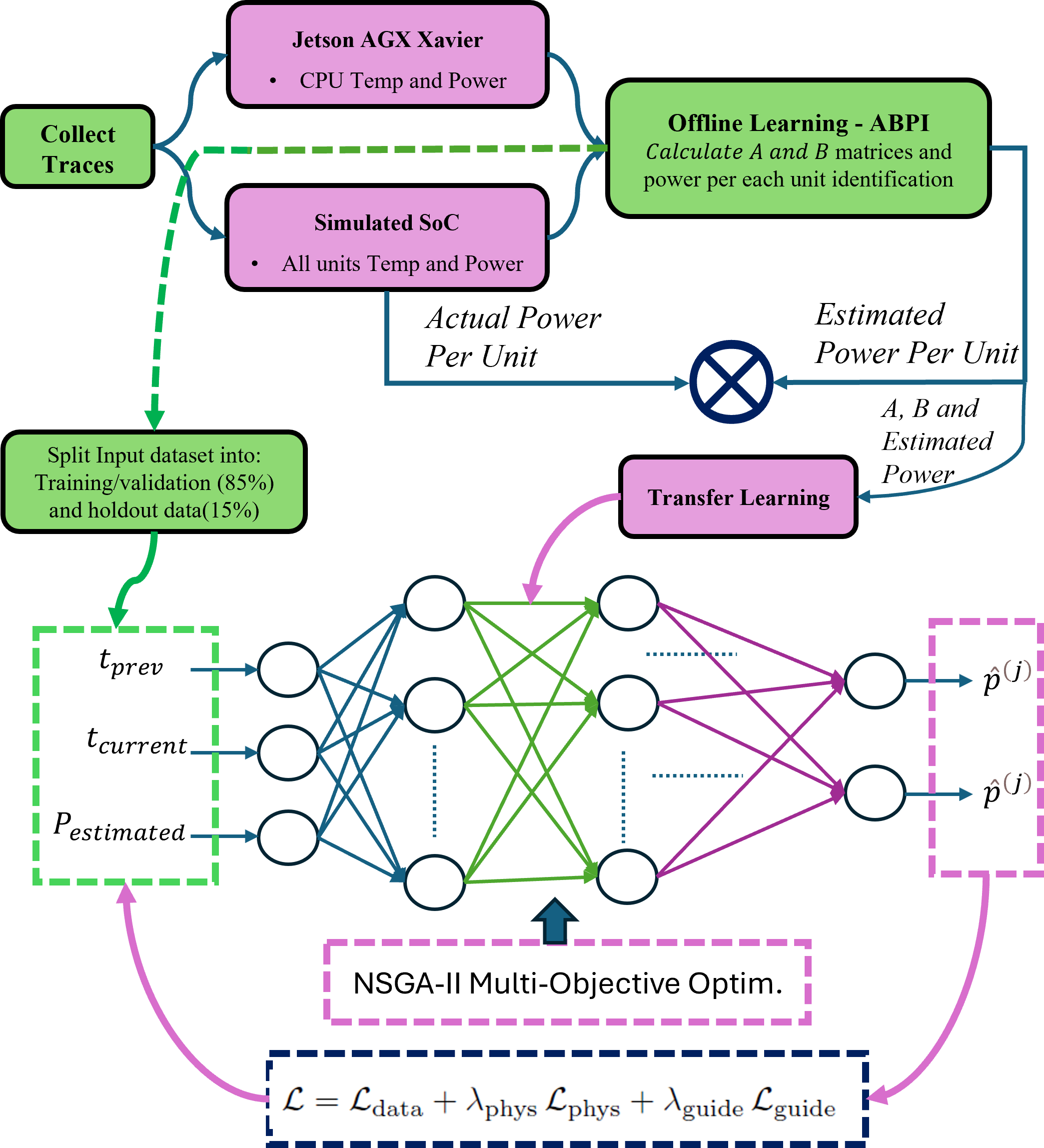}
\caption{Experimental setup overflow for the simulation and the practical validation}
\vspace{-3mm}
\label{fig:EXPOverFlow}
\end{figure}

This section details the experimental setup. \Cref{EXP_Results} will present and discuss our results. For all experiments, the ambient temperature specified in the configuration files is $298.15 K$, and the interval between the sample datasets is 1 second, including the validation experiments. Figure~\ref{fig:EXPOverFlow} is an overview of our experimental workflow, including both the simulation and empirical validation; the following describes each step in details.  

\subsection{CPINN-ABPI on NVIDIA Jetson AGX Xavier} 
\begin{figure}[!t]
    \centering
    \includegraphics[width=0.5\linewidth]{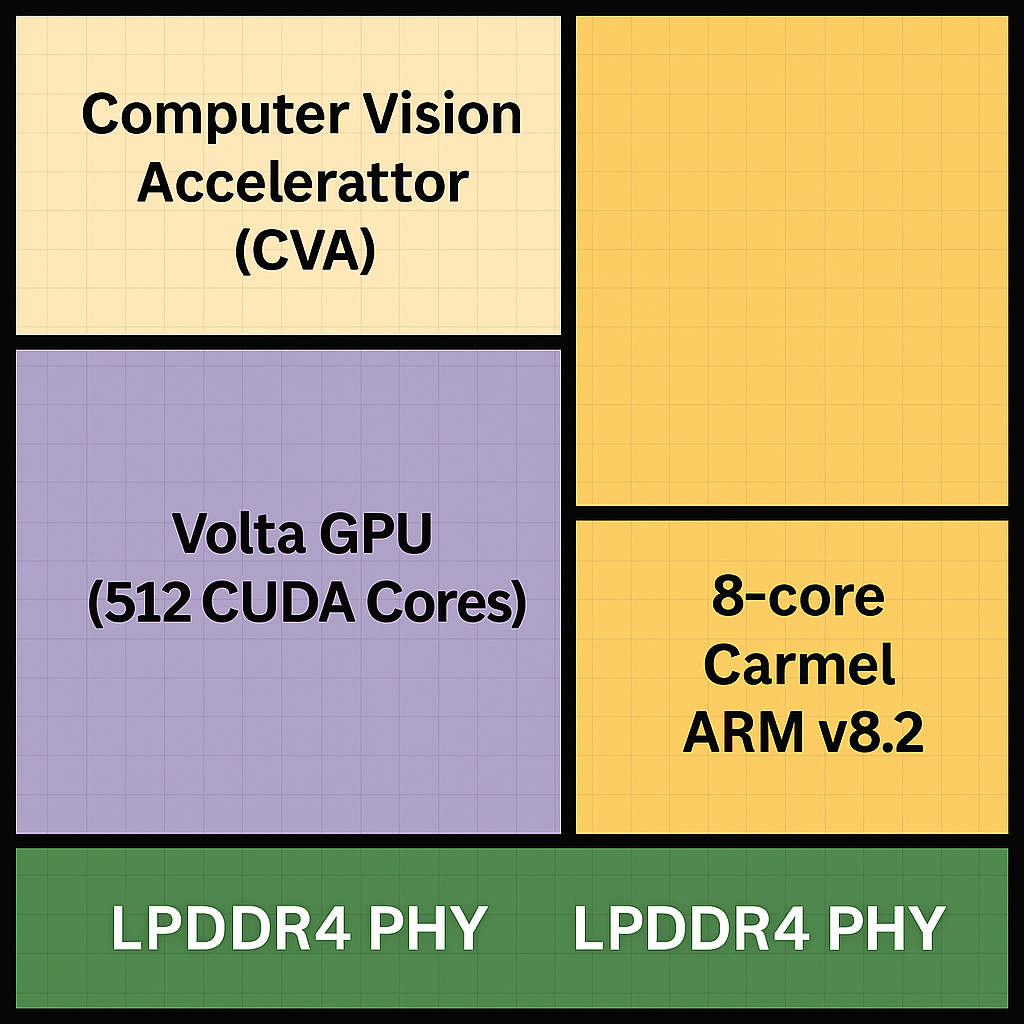}
    \caption{Simplified Jetson AGX Xavier Floorplan}
    \label{fig:JetsonFloorplan}
\end{figure}

\begin{figure}[tb]
    \centering
    \includegraphics[width=0.99\linewidth]{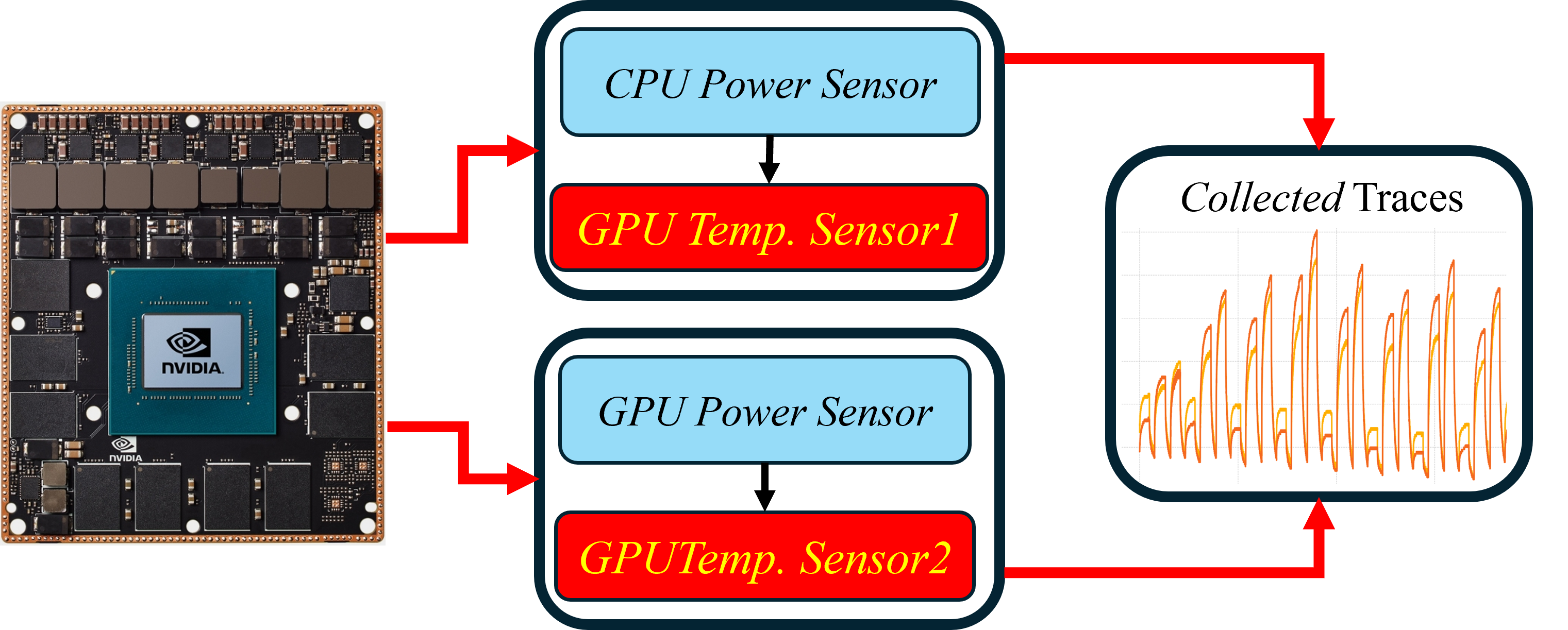}
    \caption{Jetson Xavier AGX with CPU and GPU Power and Temperature Sensors}
    \label{fig:CollectingTraces}
\end{figure}

\begin{figure}[tb]
\centering
\includegraphics[width=0.99\linewidth]{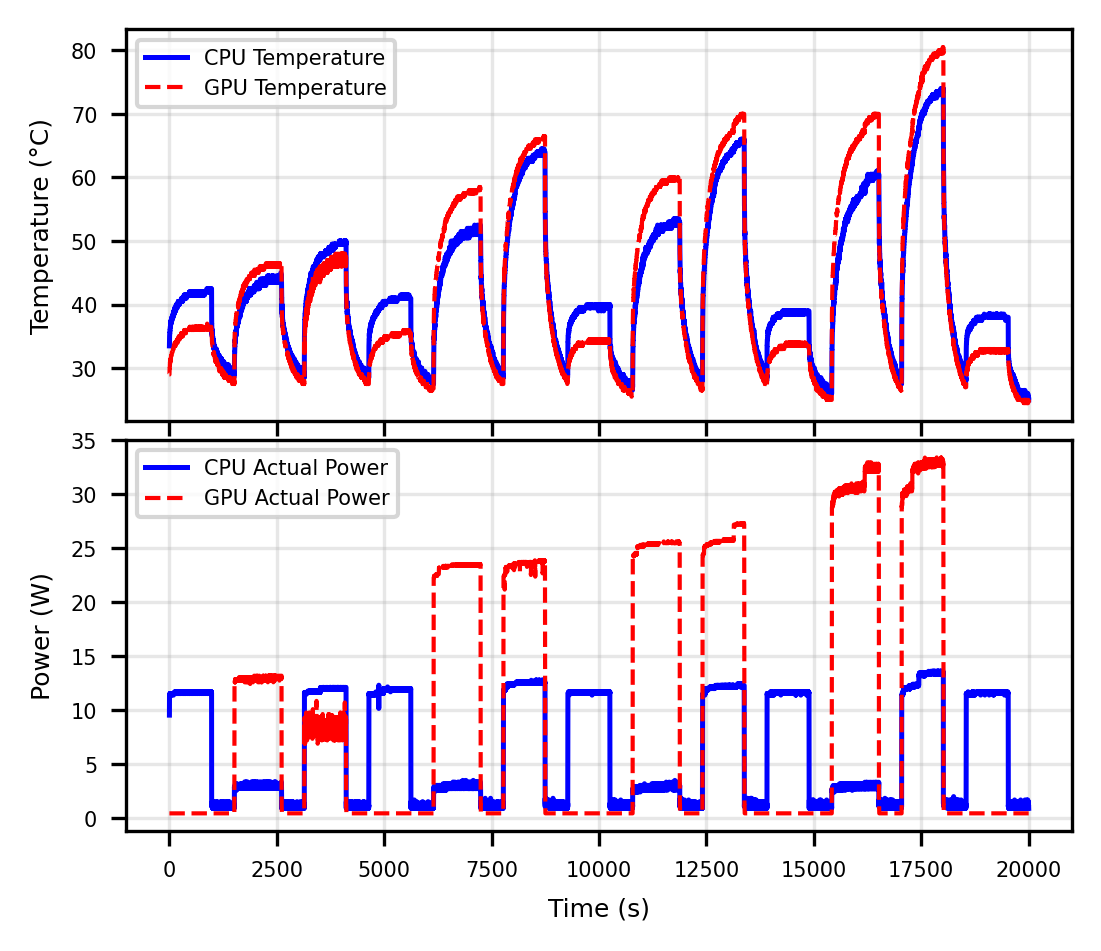}
\vspace{-6mm}
\caption{Variation in CPU and GPU temperature and Power over Time}
\label{fig:ThermalTraces1}
\end{figure}

The simplified floorplan of the Jetson AGX Xavier is shown in Figure~\ref{fig:JetsonFloorplan}. The Xavier SoC integrates dedicated power sensors for each processing unit. The Jetson features five key components: an 8-core Carmel ARM v8.2 64-bit CPU with 8MB L3 cache; an NVIDIA Volta GPU containing 512 CUDA cores and 64 Tensor Cores supporting FP16, INT8, and FP32  precision; two NVIDIA Deep Learning Accelerators (NVDLA) specialized for INT8 and FP16 inference; and a Computer Vision Accelerator (CVA) for image processing tasks. The integrated power and temperature sensors for the CPU and GPU units allow for accurate monitoring, with each component having distinct power sensors that provide detailed performance insights. Before collecting the traces, we configure the Xavier board for consistent performance by setting it to the maximum performance mode, enabling Dynamic Voltage and Frequency Scaling (DVFS), and fixing the fan speed at a duty cycle 50\%. We then apply stress and idle workloads to both the CPU and the GPU: the CPU is stressed using the \texttt{stress-ng} benchmark~\cite{king_stress_ng}, while the GPU executes continuous tensor and matrix operations. These workloads are detailed in~\Cref{table:CPU_GPU_Workloads}. Throughout the experiments, we monitor utilization with the \texttt{Nvidia Jetson Stats} tool~\cite{nvidia_jtop} to validate workload behavior. Traces are collected as shown in Figure~\ref{fig:CollectingTraces}, and thermal and power profiles are visualized in Figure~\ref{fig:ThermalTraces1}. The resulting thermal traces are then used to evaluate the performance of ABPI and CPINN-ABPI. 

\begin{table}[!t]
\centering
\caption{Workloads and Tensor Sizes for CPU and GPU (Tests 1–10)}
\label{table:CPU_GPU_Workloads}
\renewcommand{\arraystretch}{1.1} 
\setlength{\tabcolsep}{15pt} 
\begin{tabular}{c l l}
    \hline
    \textbf{Test} & \textbf{CPU Benchmark} & \textbf{GPU Tensor Size} \\ \hline
    test1  & fft         & (1024, 1024) \\
    test2  & fibonacci   & (4096, 4096) \\
    test3  & loop        & (256, 256)   \\
    test4  & loop        & (2048, 2048) \\
    test5  & matrixprod  & (512, 512)   \\
    test6  & fft         & (4096, 4096) \\
    test7  & int64       & (256, 256)   \\
    test8  & matrixprod  & (512, 512)   \\
    test9  & fibonacci   & (1024, 1024) \\
    test10 & fft         & (2048, 2048) \\
    \hline
\end{tabular}
\end{table}

\subsection{CPINN-ABPI on Heterogeneous SoC}
\begin{figure}[tb]
\centering
\includegraphics[width=0.9\linewidth]{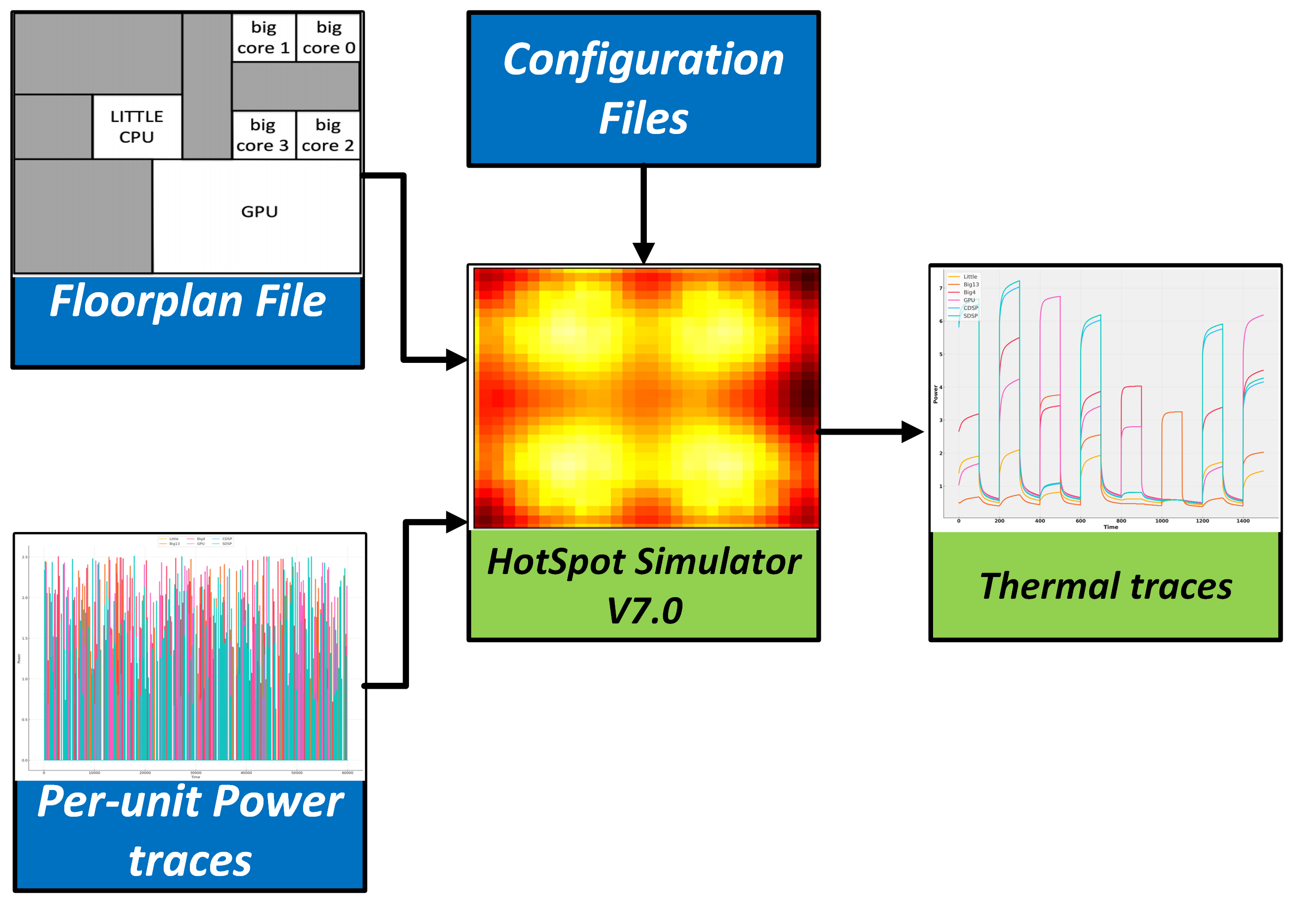}
\caption{Custom Dataset Generation Workflow for a Heterogeneous SoC}
\vspace{-3mm}
\label{fig:DataSetGenSimu}
\end{figure}

The heterogeneous SoC has an ARM big.LITTLE architecture and a dedicated GPU~\cite{Hetro_FloorPlan}. We used the HotSpot v7 thermal simulator~\cite{HotSpot} to model the temperature profiles in a heterogeneous floor plan. The floorplan has a dimension of \(1 \times 1~\text{cm}^2\) and a total power budget of 15~W. As illustrated in Figure~\ref{fig:DataSetGenSimu}, the power traces for the floorplan, along with its structural layout and the associated HotSpot configuration files, are provided as input to the simulator. The resulting thermal traces are then used to evaluate the performance of ABPI and CPINN-ABPI. 


\subsection{Unified Evaluation of ABPI and CPINN-ABPI} 
Here, we outline our evaluation of ABPI and CPINN-ABPI using simulation and real hardware data. After collecting the thermal traces from HotSpot and the Jetson Xavier board, each dataset is processed to align the power and temperature samples in a uniform 1-second interval. Initially, ABPI is applied independently to each platform to compute the thermal matrices \( A \) and \( B \) for the Jetson board and the simulated heterogeneous SoC. Using these parameters, the estimated power consumption for each unit is calculated and compared with the ground truth power measurements obtained from the onboard sensors in the real hardware and from the CoMeT simulator in the simulation environment~\cite{CoMet}. Subsequently, the proposed CPINN-ABPI approach is applied to both SoCs. The power predictions from CPINN-ABPI are then compared with those of ABPI to assess improvements in estimation accuracy, validating the effectiveness of the proposed method in both real and simulated scenarios.

\section{Experimental Results}
\label{EXP_Results}

We present a comprehensive evaluation of the CPINN-ABPI approach against ABPI. First, we detail the results of the NSGA-II multi-objective optimization used to determine the optimal neural network architecture. Then, we analyze the performance separately on the  Jetson AGX Xavier board and the simulated heterogeneous SoC. 

\subsection{NSGA-II Optimization Results}

\begin{figure*}[tb]
    \centering
    \includegraphics[width=0.99\linewidth]{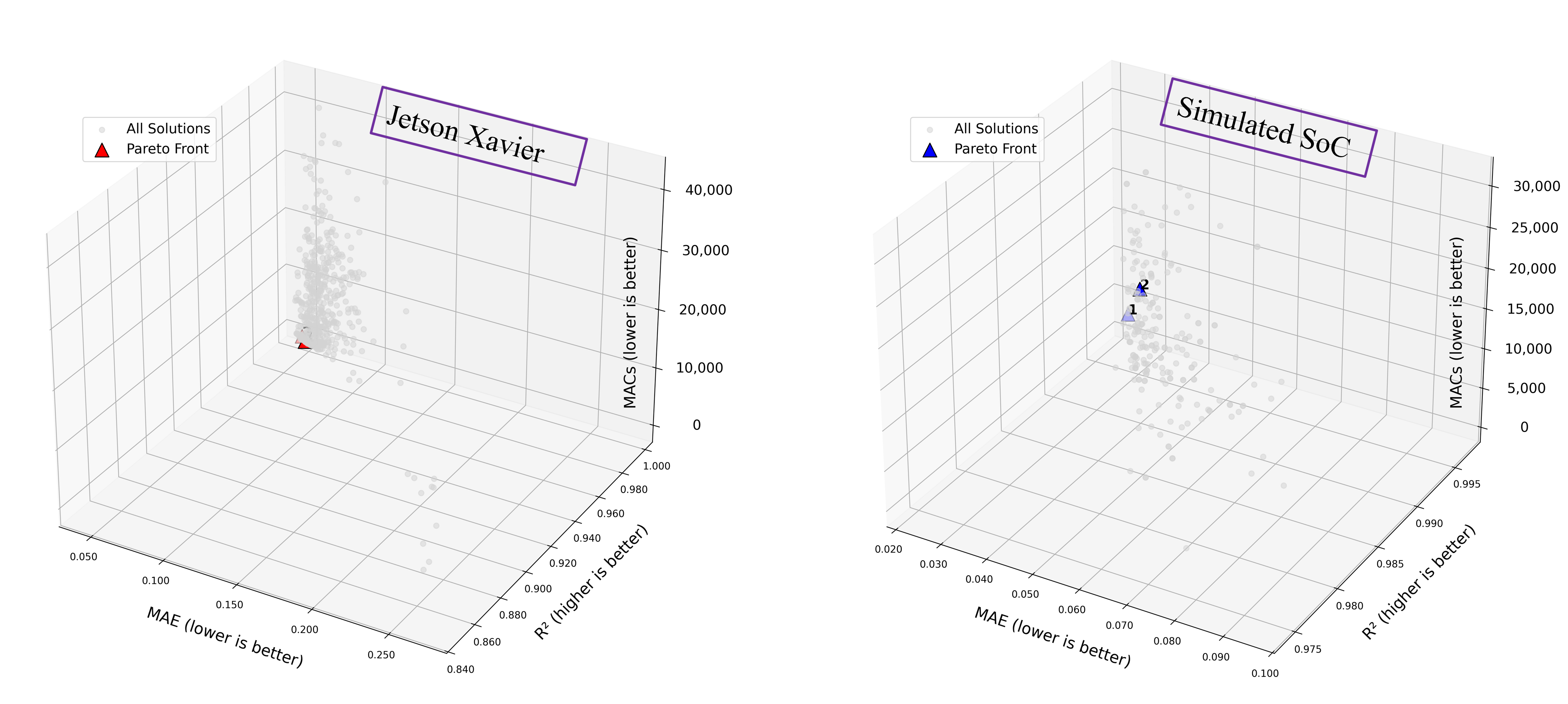}
    \vspace{-4mm}
    \caption{Pareto front showing the tradeoff between prediction accuracy (MAE in Watts) and computational cost (MAC operations) for various CPINN-ABPI configurations. The selected model configuration (marked in red) balances accuracy and efficiency.}    
    \label{fig:pareto}
\end{figure*}

\begin{figure}[t]
    \centering
    \includegraphics[width=0.99\linewidth]{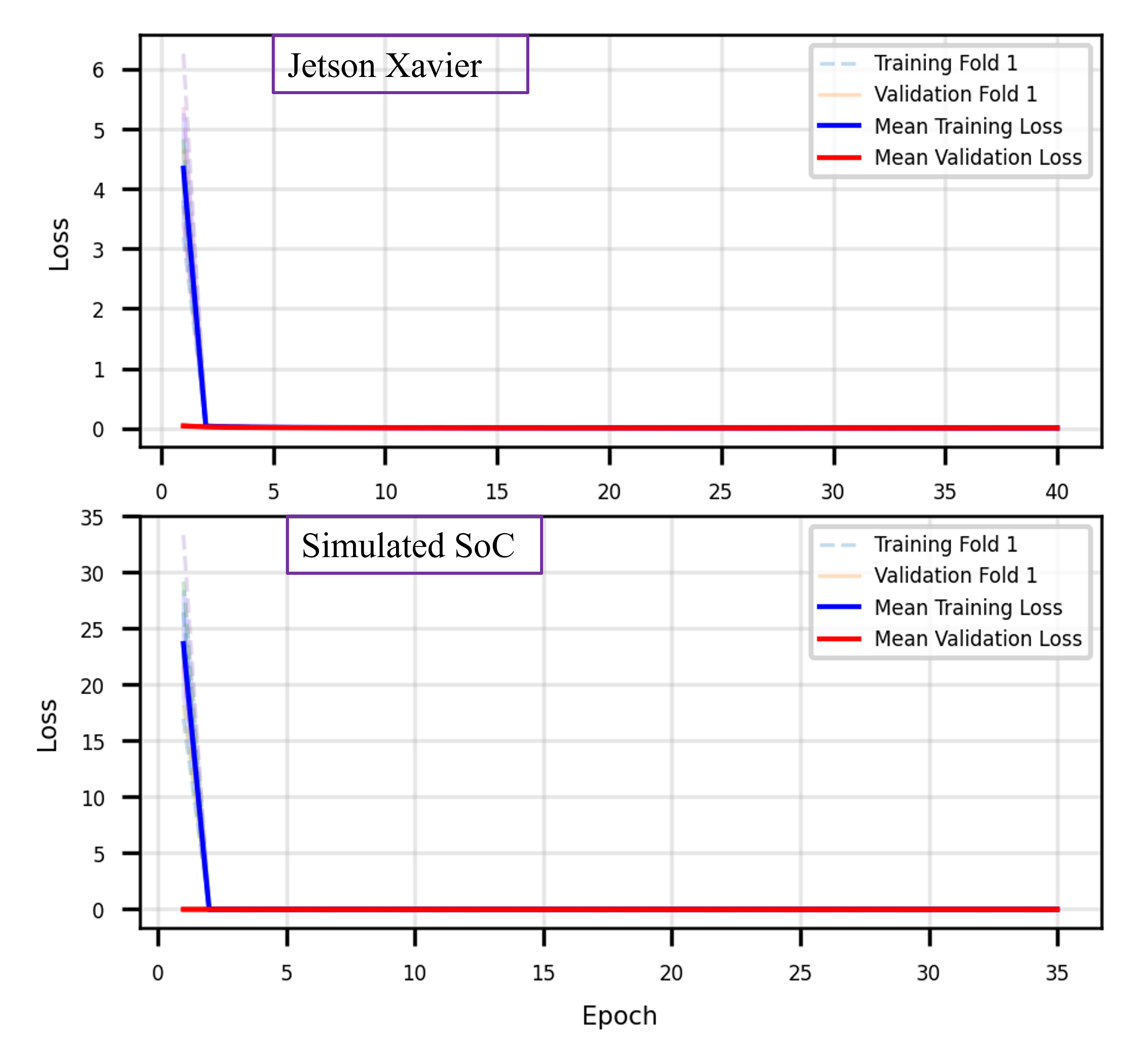}
    \vspace{-7mm}
    \caption{Average training and validation loss curves during 10-fold cross-validation for the Pareto-selected model. The blue line represents the training loss while the red line shows the validation loss across epochs. The convergence of both curves and their similar behavior indicate that the model generalizes well without significant overfitting.}
    \label{fig:cv_loss}
\end{figure}

Figure~\ref{fig:pareto} illustrates the Pareto front obtained after running NSGA-II optimization for 30 generations with a population size of 20 individuals. The optimization process explored different neural network architectures, loss function weights ($\lambda_{phys}$,$\lambda_{guide}$), and regularization parameters. The Pareto front shows the tradeoff between prediction accuracy and computational cost. We selected a 1-layer residual network with 21 neurons for the Jetson Xavier board, achieving good accuracy with only 176 MAC operations per inference for real-time power management. For the simulated SoC, we chose a 2-layer network with 80 and 55 neurons, respectively. 
To ensure that the Pareto-selected model does not overfit, we draw the average training and validation loss during the 10-fold cross-validation, as shown in Figure~\ref{fig:cv_loss}.

\begin{figure*}[t]
    \centering
    \includegraphics[width=0.99\linewidth]{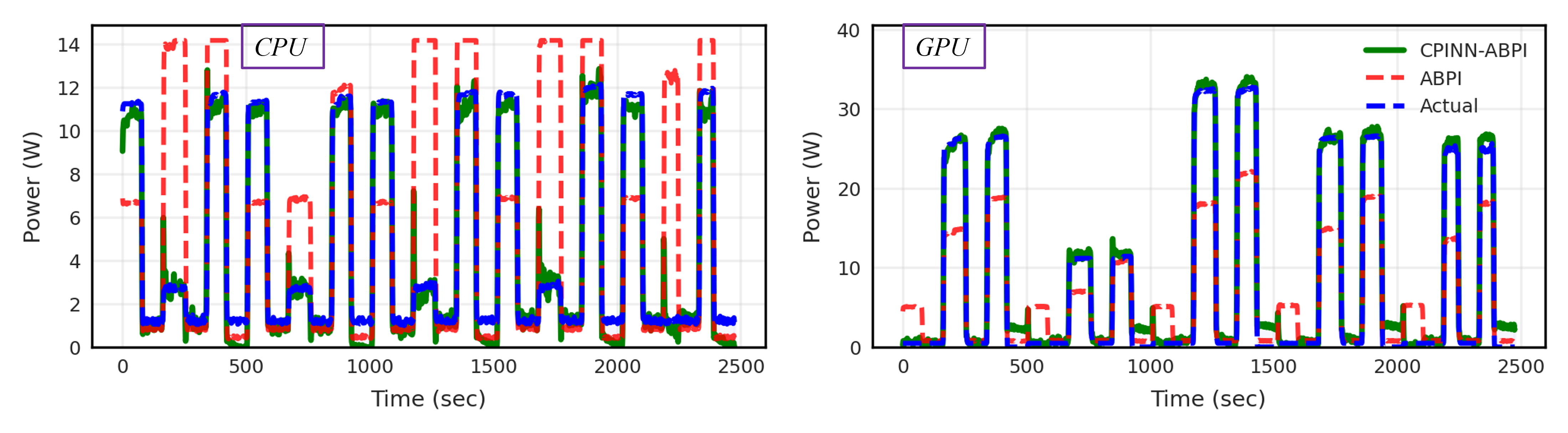}
    \vspace{-5mm}
    \caption{Comparison of actual power measurements versus predictions from ABPI and CPINN-ABPI for the CPU and GPU of the Jetson AGX Xavier under various workloads. CPINN-ABPI demonstrates significantly improved tracking of power dynamics.}
    \label{fig:xavier_prediction}
\end{figure*}

\subsection{Results on the NVIDIA Jetson AGX Xavier Board}

Figure~\ref{fig:xavier_prediction} presents a comprehensive comparison of the accuracy of power prediction among the proposed CPINN-ABPI, the baseline ABPI, and the actual measured power consumption for the CPU and GPU components over a 2500-second test dataset. CPINN-ABPI (green dashed line) demonstrates exceptional tracking accuracy, closely following the actual power traces (blue solid line) across various dynamic workload scenarios. In contrast, the baseline ABPI (red dashed line) exhibits significant prediction errors with consistently higher and more volatile estimates.

\begin{table}[tb]
\centering
\caption{Prediction Performance Metrics by Component}
\label{tab:component_performance1}
\adjustbox{width=\columnwidth,center}
{%
\begin{tabular}{lccccc}
\hline
\textbf{Method} & \textbf{Comp.} & \textbf{MAE} & \textbf{MSE} & \textbf{WMAPE} & \textbf{Time} \\
 & & \textbf{(W)} & \textbf{(W\textsuperscript{2})} & \textbf{(\%)} & \textbf{($\mu$s)} \\
\hline
\multirow{2}{*}{ABPI} & CPU & 3.74 & 32.10 & 81.02 & \multirow{2}{*}{67.2} \\
 & GPU & 3.74 & 32.10 & 47.33 & \\
\hline
\multirow{2}{*}{CPINN-ABPI} & CPU & 0.57 & 1.04 & 12.25 & \multirow{2}{*}{195.3} \\
 & GPU & 0.98 & 2.45 & 12.34 & \\
\hline
\end{tabular}%
}
\end{table}

Table~\ref{tab:component_performance1} demonstrates the superior performance of CPINN-ABPI over the baseline ABPI. CPINN-ABPI achieves substantial improvements with MAE reductions of 84.7\% (CPU) and 73.9\% (GPU), while MSE reductions exceed 96\% (CPU) and 92\% (GPU). The WMAPE values are consistently low (~12\%) for CPINN-ABPI, compared to ABPI (47--81)\%, validating the effectiveness of utilizing a PINN for accurate power prediction. Regarding computational efficiency, CPINN-ABPI introduces a moderate overhead with inference times of 195.3 $\mu$s compared to  67.2 $\mu$s (ABPI) on an NVIDIA A100 GPU, representing a 2.9× increase in computational cost. However, both methods maintain sub-millisecond inference capabilities suitable for real-time applications, while the dramatic accuracy improvements (over 90\% MSE reduction) justify the computational tradeoff for high-precision power monitoring scenarios.

\subsection{Results on a Simulated Heterogeneous SoC} 

\begin{figure*}[tb]
    \centering
    \includegraphics[width=0.99\textwidth]{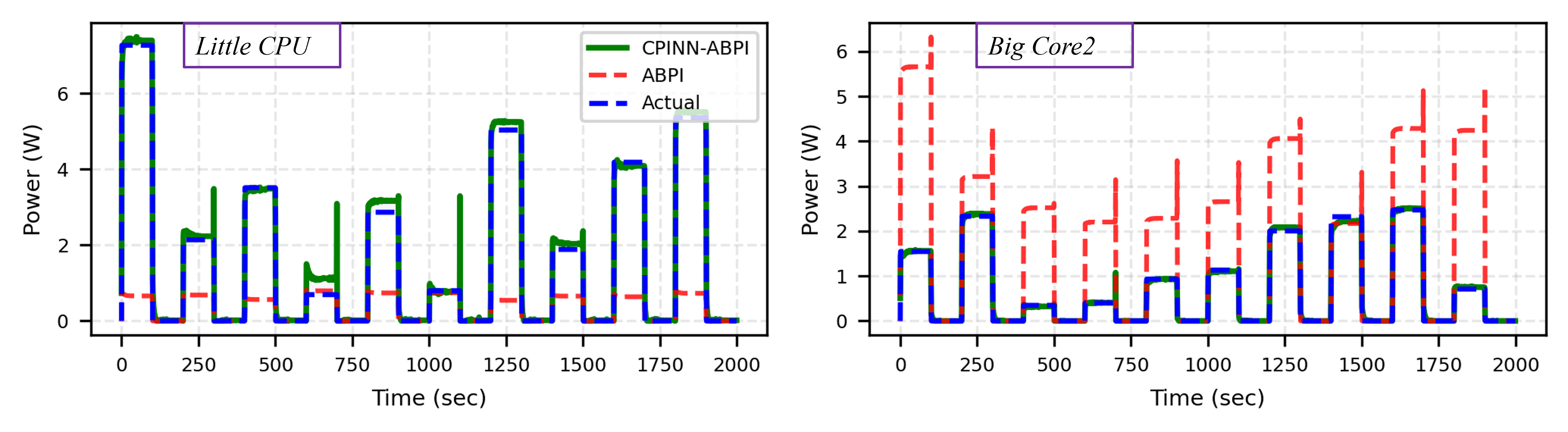}
    \vspace{-3mm}
    \caption{Power prediction comparison between CPINN-ABPI and ABPI. (Left) Little CPU unit performance over 2000 seconds. (Right) Big Core2 performance over the same period. The proposed CPINN-ABPI method demonstrates superior accuracy compared to the baseline ABPI.} 
    \label{fig:power_prediction_comparison}
\end{figure*}

Figure~\ref{fig:power_prediction_comparison} shows the power prediction results for our simulated heterogeneous SoC with a big.LITTLE architecture. The results demonstrate that CPINN-ABPI consistently outperforms ABPI across all SoC components, with notable improvements in GPU power prediction accuracy. Performance metrics demonstrate the substantial effectiveness of CPINN-ABPI in the heterogeneous six-component architecture. Table~\ref{tab:component_performance} shows that all processing units, from the Little CPU to the four Big Cores and GPU, exhibit dramatic MAE improvements between 85--99\% when transitioning from baseline ABPI to CPINN-ABPI. In particular, the GPU and Big Core 3 achieve exceptional accuracy with MAE values of only 0.01~W, representing the best performance across the heterogeneous system. Although ABPI struggles with highly variable WMAPE values (ranging from 38.7--570.2\%), indicating inconsistent relative accuracy across components, CPINN-ABPI demonstrates more stable performance characteristics. The computational overhead of 84.6\%, going from 141.9~$\mu$s to 261.9~$\mu$s per inference, proves justified given the substantial accuracy gains, with both methods maintaining real-time capability at over 3,800 inferences per second. This performance validates the effectiveness of CPINN-ABPI in handling the diverse power characteristics inherent in heterogeneous computing architectures.

\begin{table}[tb]
\centering
\caption{Prediction Performance Metrics by Component}
\label{tab:component_performance}
\adjustbox{width=\columnwidth,center}
{%
\begin{tabular}{lcccc}
\hline
\textbf{Method} & \textbf{Comp.} & \textbf{MAE} & \textbf{WMAPE} & \textbf{Time} \\
 & & \textbf{(W)} & \textbf{(\%)} & \textbf{($\mu$s)} \\
\hline
\multirow{6}{*}{ABPI} & Little CPU & 1.66 & 38.7 & \multirow{6}{*}{141.9} \\
 & Big Core 0 & 0.37 & 262.4 & \\
 & Big Core 1 & 0.30 & 121.7 & \\
 & Big Core 2 & 1.25 & 570.2 & \\
 & Big Core 3 & 0.42 & 80.3 & \\
 & GPU & 1.03 & 344.4 & \\
\hline
\multirow{6}{*}{CPINN-ABPI} & Little CPU & 0.11 & 50.2 & \multirow{6}{*}{261.9} \\
 & Big Core 0 & 0.03 & 125.9 & \\
 & Big Core 1 & 0.04 & 18.7 & \\
 & Big Core 2 & 0.05 & 31.7 & \\
 & Big Core 3 & 0.01 & 16.8 & \\
 & GPU & 0.01 & 8.1 & \\
\hline
\end{tabular}%
}
\end{table}

\section{Conclusion} 
\label{sec:conclusion}

This paper presents the first empirical validation of Alternative Blind Power Identification (ABPI) on a commercial MPSoC hardware, revealing that despite its theoretical appeal, ABPI suffers from significant accuracy limitations in real-world scenarios. To overcome these shortcomings, we introduce CPINN-ABPI, a novel, hybrid power estimation approach that integrates physics-based models with data-driven learning through Custom Physics-Informed Neural Networks. Our results demonstrate that CPINN-ABPI achieves substantial accuracy gains across diverse platforms, reducing power estimation errors by over $80\%$ while maintaining sub-millisecond inference times. This work is a critical advancement for real-time power monitoring in heterogeneous multicore systems. The proposed approach will help achieve more reliable thermal and power management, enhancing dynamic voltage and frequency scaling, bettering detection of thermal attacks, and improving energy efficiency in data centers. Its flexibility and precision position CPINN-ABPI as a foundational approach for next-generation system architectures, where accurate, fine-grained power estimation is essential. 



\newpage
\bibliographystyle{IEEEtran}
\bibliography{References} 

\end{document}